\documentclass[twocolumn,twoside,slac_two]{revtex4}
\usepackage{graphicx}
\usepackage{fancyhdr}
\def\thc {3C~273}

\begin{document}

\title{The variability of the quasar \thc: a radio to gamma-ray view}

\author{Simona Soldi}
\affiliation{Laboratoire AIM - CNRS - CEA/DSM - Universit\'e Paris Diderot (UMR 7158), CEA Saclay, 
	     DSM/IRFU/SAp, F91191 Gif-sur-Yvette, France}

\author{Volker Beckmann}
\affiliation{APC, Centre Fran\c{c}ois Arago, IN2P3/CNRS - Universit\'e Paris Diderot, 
	     10 rue Alice Domon et L\'eonie Duquet, 75205 Paris Cedex 13, France}

\author{Marc T\"urler}
\affiliation{ISDC Data Centre for Astrophysics, Observatory of the University of Geneva, Chemin d'Ecogia 16, Versoix, Switzerland}

\begin{abstract}
We have analysed the first 15 months of \textit{Fermi}/LAT data of the radio loud quasar \thc.
Intense gamma-ray activity has been detected, showing an average flux of $F(> 100 \, \rm MeV) = 1.4 \times 10^{-6} \, \rm ph \, cm^{-2} \, s^{-1}$, 
with a peak at $F(> 100 \, \rm MeV) = 5.6 \times 10^{-6} \, \rm ph \, cm^{-2} \, s^{-1}$ detected during a flare in September 2009. 
Together with the brightening of the source, a possible
hardening of the gamma-ray spectrum is observed, pointing to a shift of the inverse Compton peak toward
higher energies than the 1--10 MeV range in which \thc\ inverse Compton emission is typically observed to peak.
During the 15 months of observations the photon index is measured to vary between $\Gamma = 2.4$ and $\Gamma = 3.3$, with an average value
of $\langle \Gamma \rangle = 2.78 \pm 0.03$.
When compared to the observations at other wavelengths, the gamma-rays show the largest flux variations
and we discuss the possibility that two different components are responsible for the inverse Compton hump
emission below and above the MeV peak.
\end{abstract}

\maketitle

\thispagestyle{fancy}


\section{Introduction}

\thc\ is a bright, radio loud quasar that shares most of the characteristic properties of blazars, like strong radio emission,
a jet with apparent superluminal motion, large flux variations, and, occasionally, polarisation of the optical emission 
(see Courvoisier et al. 1998 for a review). On the other hand, this AGN presents also characteristics more common in Seyfert galaxies,
like a prominent blue bump \cite{shields78} and the signature of non-thermal Comptonisation processes in the X-ray spectrum \cite{grandi04}.

\thc\ was the first extragalactic source to be detected at gamma-rays by \textit{COS-B} observations in 1976 \cite{swanenburg78} and later on
was detected several times by the EGRET instrument on board the \textit{CGRO} satellite \cite{vonmontigny97}.
Thanks to its early discovery as a multiwavelength emitter and to its brightness, \thc\ is one of the best observed AGN at all energies.
The multiwavelength on-line database of \thc\ (\emph{http://isdc.unige.ch/3c273/}) contains about 40 years of data from radio to gamma-rays \cite{soldi08}.
This set of data constitutes a very precious starting point for studying the multi-band variability of this object. 
In particular, with the launch of \textit{Fermi} in June 2008 an intense gamma-ray monitoring of the source has started \cite{abdo09a} and triggered observations 
along the whole electromagnetic spectrum, coordinated by several groups. This extensive and regular look at the source, together with
the comparison with the historical long-term behaviour at all wavelengths, will allow us to investigate the properties and origin of the high-energy emission
in greater detail than it had ever been done before.

\section{From radio to X-rays}

\subsection{The \thc\ multiwavelength database}

\begin{figure}[b]
\centering
\hspace{-0.5cm}
\includegraphics[width=85mm]{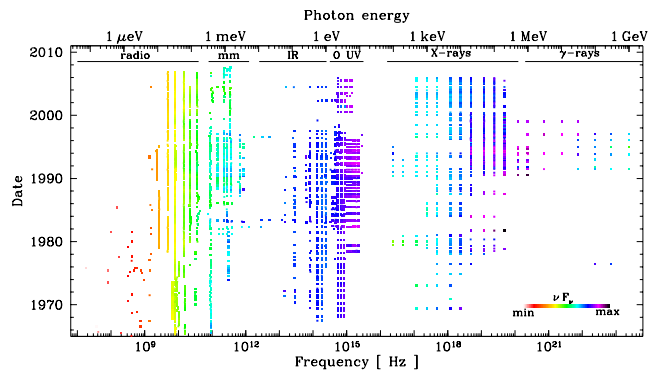}
\caption{Time coverage of all \thc\ observations of the database as a function of the frequency, color-coded to represent the $\nu \rm F_{\nu}$ 
         intensity of the measured fluxes.} 
\label{database}
\end{figure}

The \thc\ database, first created in 1999 \cite{turler99}, has been recently updated with observations up to 2007 \cite{soldi08} in order to provide 
the community with one of the most complete multiwavelength databases for an AGN.
At the time being the database contains 70 light curves covering 17 decades in energy and spanning up to 40 years of observations (Fig.~\ref{database}).
In addition, for the X-ray band, information about the spectral parameters used to extract the fluxes is also available.
The bulk of the high-energy data has been collected in the last 10--15 years, thanks to the intense X-ray monitoring provided by \textit{BeppoSAX},
\textit{CGRO}/BATSE, \textit{RXTE}, \textit{XMM-Newton}, \textit{INTEGRAL} and \textit{Swift}. In the gamma-ray domain, the data collected by
\textit{CGRO}/OSSE and EGRET have already been included in the database, while we plan to include also results from \textit{AGILE}/GRID and \textit{Fermi}/LAT 
observations, within the upcoming months.

\subsection{The variability study}
As a first application of the on-line database, we studied the variability properties of \thc\ at different wavelengths and, correlating
the behaviour in the different energy bands, we tried to identify the emission processes and understand their inter-connections \cite{soldi08}.

\begin{figure}[t]
\centering
\includegraphics[width=60mm,angle=90]{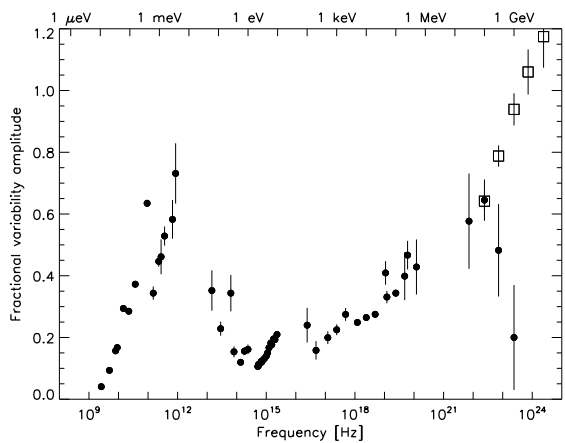}
\caption{Spectrum of the fractional variability amplitude, computed using the light curves of the \thc\ database. 
	 Above 100~MeV, the circles represent the variability observed with EGRET data and the open squares are the results obtained
	 by the LAT.}
\label{fvar}
\end{figure}

The amplitude and the time scales of the variations of \thc\ depend strongly on the frequency and show trends that are characteristic 
of the underlying emission processes (Fig.~\ref{fvar}). For example, the increasing amplitude of the variations and the decreasing
time scales when going from long radio to short millimeter wavelengths is the typical signature of the superposition of synchrotron flares from
shock waves propagating along the jet.

The different variability properties in the X-rays below and above 20 keV seem to support the possibility of two separated
emitting components, possibly non thermal Comptonisation (as in Seyfert galaxies) and inverse Compton from the jet (as in blazars).
As an alternative, a single component characterised by at least two independently varying parameters could also explain
the X-ray timing properties.

\begin{figure}[t]
\centering
\includegraphics[width=60mm,angle=90]{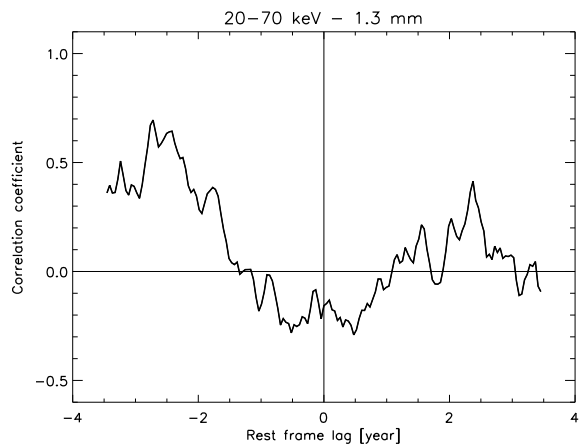}
\caption{Correlation between the hard X-ray (20--70 keV) and the 1.3 mm light curves. No correlation is detectable at short time lags.} 
\label{cross}
\end{figure}

In any case, the dominant hard X-ray ($> 20 \, \rm keV$) emission is most probably not due to electrons accelerated by 
the shock waves in the jet as their variability does not correlate with the flaring millimeter emission at short time lags (Fig.~\ref{cross}). 
Instead, the hard X-rays seem to be associated to long-timescale variations in the optical domain. This optical component is 
consistent with being optically thin synchrotron radiation from the base of the jet and the hard X-rays would be produced through 
inverse Compton processes (synchrotron self-Compton and/or external Compton) by the same electron population. 

In the gamma-ray domain, the limited data available before \textit{AGILE} and \textit{Fermi} observations did not allow to draw 
a firm conclusion about the variability in the MeV-GeV energy range. The increase of the variability amplitude observed in the X-rays
seems to continue up to the gamma-rays, possibly even above 100~MeV where the observed decrease is likely due to the large
uncertainties of the EGRET fluxes (circles in Fig.~\ref{fvar}).

\section{The gamma-ray emission}

\subsection{Before \textit{Fermi}}
During \textit{CGRO}/EGRET observations, \thc\ was either found below the EGRET sensitivity level, or detected at a relatively 
low gamma-ray flux, with an average flux of $\rm F(> 100 \, \rm MeV) = 15 \times 10^{-8} \, \rm ph \, cm^{-2} \, s^{-1}$, between 1991 and 1995, and
a maximum detected flux of $\rm F(> 100 \, \rm MeV) \simeq 80 \times 10^{-8} \, \rm ph \, cm^{-2} \, s^{-1}$ in 1997 \cite{collmar00}.
The EGRET average spectrum has a power law shape with photon index $\langle \Gamma \rangle = 2.4 \pm 0.1$, but variations of the spectral slope have been
observed, with the spectrum hardening while the source brightened \cite{vonmontigny97,collmar00}.
A flux variability by a factor of 4 within one week is the most extreme behaviour observed for \thc\ at gamma-rays by EGRET \cite{collmar00}.
Multiwavelength campaigns have been carried out simultaneously to the the gamma-ray observations in 1993--1995 and showed that the emission detected by EGRET
can be equally well explained with synchrotron self-Compton (SSC), external Compton (EC) or proton-initiated cascade (PIC) models \cite{vonmontigny97}.

During the first year of observations, \textit{AGILE}/GRID has detected \thc\ with an average flux of 
$\rm F(> 100 \, \rm MeV) = 24 \times 10^{-8} \, \rm ph \, cm^{-2} \, s^{-1}$, consistent with EGRET measurements \cite{pittori09}.
During the multiwavelength campaign of December--January 2008, the source was detected only during the second week of observations at a flux of 
$\rm F(> 100 \, \rm MeV) = 34 \times 10^{-8} \, \rm ph \, cm^{-2} \, s^{-1}$ \cite{pacciani09}.
The hard X-ray to gamma-ray emission is modeled as external Compton radiation on thermal disc photons, and the 
variability of this high-energy component is consistent with an acceleration phase of the electron distribution that moves the inverse 
Compton peak to higher energies.

\subsection{\textit{Fermi}/LAT data analysis}

\begin{figure}[t]
\centering
\includegraphics[width=70mm]{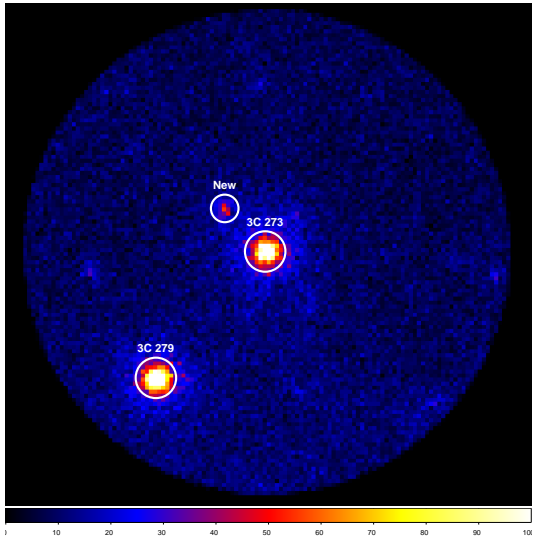}
\caption{\textit{Fermi}/LAT count map of the 15$^\circ$ region around \thc, produced with data collected during 15 months. 
	 Two additional sources are clearly detected: the blazar 3C~279 and the unidentified source at $\rm R.A. = 189.9$, $\rm DEC = 4.7$.} 
\label{Fermi_map}
\end{figure}

The \textit{Fermi} satellite, launched in June 2008, carries two instruments able to observe the gamma-ray sky: the Large Area Telescope (LAT),
a pair conversion gamma-ray telescope sensitive to gamma-rays from 30~MeV to 200~GeV \cite{atwood09}; and the GLAST Burst Monitor (GBM), observing the sky in the 
8 keV to 30 MeV energy range \cite{meegan09}.
The main instrument LAT is 10 times more sensitive than its predecessors \textit{CGRO}/EGRET and \textit{AGILE}/GRID and, thanks to its large field
of view of $\sim 2.4 \, \rm sr$, it can scan the full sky every 3 hours. 
Already after the first 3 months of observations, the LAT had detected 205 objects with high-confidence, 
of which 106 are AGN \cite{abdo09a,abdo09b}, including \thc.

We analysed the \textit{Fermi}/LAT data collected during the first 15 months of observations, between August 10, 2008 and October 19, 2009.
We first downloaded the 0.1--300~GeV light curve of \thc\ provided through the \textit{Fermi Monitored Source List Light Curves} 
page\footnote{\emph{http://fermi.gsfc.nasa.gov/ssc/data/access/lat/msl\_lc/}}, selected only the detections on daily basis and rebinned
the light curve to 3-day bins. We then performed the analysis on each of these 3-day time intervals selecting the event data with the \textit{gtselect}
tool distributed as part of the \textit{Fermi}-LAT ScienceTools (version 9.15.2).
For each bin (corresponding in average to an integration time slightly smaller than 200~ks), the count map and exposure map 
were generated. 

For source detection and flux estimate, we used the \textit{gtlike} tool, based on a maximum likelihood algorithm. For a region of interest (ROI) 
of 15$^\circ$ around \thc, a model containing the isotropic and Galactic diffuse background models (as given by default) and all the sources in the 
field was fitted to the data. 
For the field and data analysed, beside \thc\ two additional sources were considered: the blazar 3C~279 and the unidentified source at 
$\rm R.A. = 189.9$, $\rm DEC = 4.7$ (Fig.~\ref{Fermi_map}). All the sources have been modeled with simple power laws in the energy range 0.1--200~GeV, 
in order to obtain their spectral parameters and test statistic, TS. We note that for each source the square root of 
TS provides an approximate estimate of the detection significance.

Following \cite{abdo09a}, we added 3\% systematics to the uncertainties on the estimated fluxes.

\subsection{\textit{Fermi}/LAT preliminary results}

\begin{figure}
\centering
\includegraphics[width=80mm]{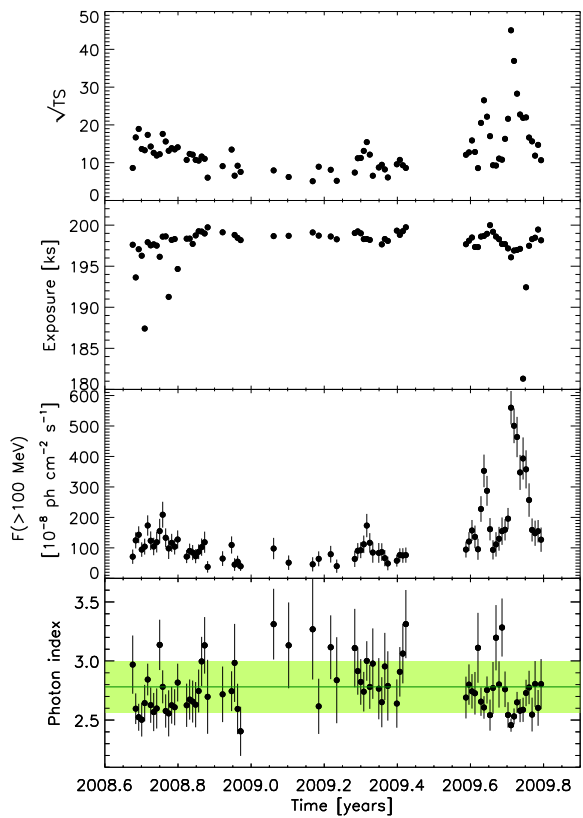}
\caption{LAT results after about one year of observations of \thc. From the top to bottom panels: square root of the test statistic (corresponding to the detection
	significance); integration time for each data bin; gamma-ray flux above 100~MeV; photon index from a simple power law fit. In the bottom panel,
	the line indicates the average photon index over 15 months and the shaded area indicates the dispersion from the mean.} 
\label{Fermi_all}
\end{figure}

Figure~\ref{Fermi_all} shows the results obtained during the first 15 months of observations. In the selected time
intervals analysed, \thc\ was always detected with $\sqrt{TS} \geq 5$. 
Since the launch of \textit{Fermi}, \thc\ has undergone several episodes of very intense gamma-ray activity.
The average flux above 100~MeV, $F(> 100 \, \rm MeV) = 1.4 \times 10^{-6} \, \rm ph \, cm^{-2} \, s^{-1} = 5.2 \times 10^{-10} \, \rm erg \, cm^{-2} \, s^{-1}$, 
is significantly larger than the historical measurements of EGRET, with a peak flux of $F(> 100 \, \rm MeV) = 5.6 \times 10^{-6} \, \rm ph \, cm^{-2} \, s^{-1}$ 
detected during a large flare in September 2009.
The amplitude of the variations, $F_{\rm var}$ (quantified through the fractional variability amplitude; see \cite{soldi08} for more detail),
measured by \textit{Fermi}, is consistent with the EGRET one at 100~MeV, but it is significantly larger above 100~MeV, confirming the
increasing trend of $F_{\rm var}$ already observed in the X-ray domain (Fig.~\ref{fvar}). The decrease of $F_{\rm var}$ observed with EGRET
above 100~MeV was likely due to its limited sensitivity. 
In addition, the amplitude of the variations measured in the gamma-rays by \textit{Fermi} is the highest across the whole electromagnetic spectrum,
confirming the extreme variability nature of the gamma-ray emission.

\begin{figure}[t]
\centering
\includegraphics[width=75mm]{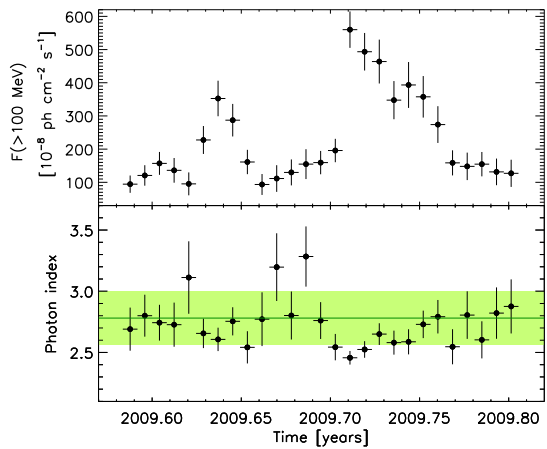}
\caption{Zoom of Figure~\ref{Fermi_all} on the last 3 months of observations. During the large flare of September 2009, the spectrum of the source seems
	 to be significantly harder then outside the flaring period.} 
\label{Fermi_flare}
\end{figure}

\begin{figure}[b]
\centering
\includegraphics[width=60mm,angle=90]{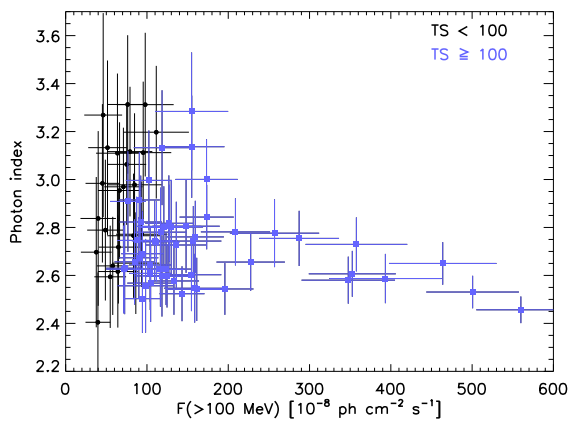}
\caption{LAT gamma-ray photon index versus gamma-ray flux of \thc. Blue (black) points represent detections with TS larger (smaller) than 100.
	The source behaviour ``harder when brighter'' is observed in the tail at fluxes above $2 \times 10^{-6} \, \rm ph \, cm^{-2} \, s^{-1}$
	with photon indices always below the average value of $\Gamma = 2.8$.} 
\label{Fermi_scatter}
\end{figure}

The bottom panel of Figure~\ref{Fermi_all} shows the evolution of the gamma-ray photon index during the LAT observations.
The photon index varies between $\Gamma = 2.4 \pm 0.2$ and $\Gamma = 3.3 \pm 0.3$, with an average value of $\langle \Gamma \rangle = 2.78 \pm 0.03$.
In general, the variations of the spectral slope do not seem to be significant, when the uncertainties are taken into account, with the exception of a 
decrease of the photon index when the source brightened during the last 3 months of observations (Fig.~\ref{Fermi_flare}).
In fact, when plotting the photon index versus flux, the source behaviour ``harder when brighter'' is clearly observed (Fig.~\ref{Fermi_scatter}): when \thc\ was brighter
than $2 \times 10^{-6} \, \rm ph \, cm^{-2} \, s^{-1}$, the photon index was always smaller than 2.8.
The same effect was observed also in EGRET data during two flares of \thc\ \cite{vonmontigny97,collmar00} and seems to be a common 
behaviour of the gamma-ray emission of blazars (see for example 3C~454.3 and AO~0235+164; \cite{lott09,escande09}).
As proposed by Pacciani et al. (2009) to explain the gamma-ray variability observed with \textit{AGILE},
the inverse Compton peak might be shifted toward higher frequencies following an accelaration episode of the electron population. 
Since the inverse Compton peak of \thc\ usually falls in the 1--10 MeV range \cite{johnson95,vonmontigny97} (Fig.~\ref{sed}),
a shift toward higher energies could indeed produce a harder spectrum in the LAT energy band.

\section{The inverse Compton peak}

\begin{figure}[t]
\centering
\includegraphics[width=50mm,angle=90]{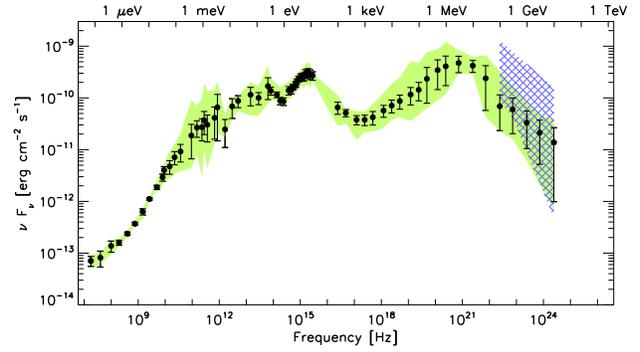}
\caption{Average spectral energy distribution of \thc\ over up to 40 years. The error bars represent the standard deviation from the mean values 
         and the green area indicates the observed range of variations. Above 30~MeV, the circles correspond to the average EGRET data, 
	 whereas the blue cross hatched area indicates the flux range of \thc\ during \textit{Fermi}/LAT observations.} 
\label{sed}
\end{figure}

Gamma-ray observations of \thc\ are essential to constrain the overall shape of the inverse Compton branch and to better understand the 
connection between the jet emission in different energy domains.
The most recent radio to gamma-ray campaigns carried out on this object, together with the numerous ones performed in the past, provide a large set 
of data to model the spectral energy distribution (SED) during several epochs and with different integration times (the average SED in Fig.~\ref{sed}).

Multiwavelength campaigns performed during EGRET observations could not discriminate between the most common models used to describe
the high-energy emission of blazars \cite{vonmontigny97}. A different variability behaviour below and above $\sim 3 \, \rm MeV$
suggested that two different mechanisms are responsible for the emission in the MeV and in the EGRET domain, with the latter
one being most likely produced by inverse Compton scattering of accretion disc photons \cite{collmar00}.
On the other hand, the different variability of the hard X-rays and gamma-rays detected by \textit{INTEGRAL} and \textit{AGILE}
has been explained by a single external Compton component shifted toward higher energies following an acceleration of the electron population,
whereas a synchrotron self-Compton contribution is evident at lower energies \cite{pacciani09}.

Several multiwavelength monitoring programs are on-going to follow the evolution of \thc\ and many other gamma-ray sources.
The preliminary results presented during the \textit{2009 Fermi Symposium} for \thc\ indicate that there is a correlation between
the gamma-ray flaring states and the brightening of the VLBI core, with gamma-ray flare following the ejection of a superluminal knot
by up to 70 days \cite{jorstad09}. This would indicate that the gamma-ray high states are related to a disturbance crossing the millimeter-jet core.
An increase of the radio emission at 15~GHz has been observed quasi-simultaneously to the large September 2009 gamma-ray flare,
and the optical flux started also to increase in July/August 2009, but no data were available at the time of the gamma-ray flare \cite{chatterjee09}.
In addition, a correlation between the small gamma-ray flare of April 2009 and the optical emission is clearly detected \cite{chatterjee09}.

As the flaring optical emission of \thc\ is believed to be synchrotron emission related to the millimeter jet, a gamma-ray/optical correlation 
would support the origin of the high-energy emission in the millimeter jet and could indicate that the same electron population is responsible for 
both emissions through synchrotron and inverse Compton processes. On the other hand, we found the hard X-ray emission to be related rather
to another optical component, varying on longer time scales, and produced as synchrotron emission further away from the millimeter jet \cite{soldi08}.
Therefore, this would point to a different origin of the hard X-ray and gamma-ray emissions.

\section{Conclusions}

Thanks to its large field of view and its unprecedented sensitivity above 100 MeV, the LAT instrument is providing us
with long-term, well sampled light curves and spectral information for a large number of AGN \cite{abdo09b,lott09,beckmann09}.
Among them, \thc\ is a bright quasar extensively observed at all wavelengths and showing a particularly intense
gamma-ray activity since the launch of \textit{Fermi}.
This source was detected during the first 15 months of observations with an average flux of 
$F(> 100 \, \rm MeV) = 1.4 \times 10^{-6} \, \rm ph \, cm^{-2} \, s^{-1}$, corresponding to an average luminostity
of $\langle L_{\gamma} \rangle = 3 \times 10^{46} \, \rm erg \, s^{-1}$, and an average photon index of $\langle \Gamma \rangle = 2.78 \pm 0.03$.
In September 2009, during a very bright flare reaching a peak flux of $F(> 100 \, \rm MeV) = 5.6 \times 10^{-6} \, \rm ph \, cm^{-2} \, s^{-1}$, 
the hint of a spectral hardening accompanied the brightening of the source, possibly indicating a shift of the inverse Compton peak 
from 1--10 MeV (as typically observed in \thc) toward higher energies.

When looking at the multi-band behaviour, the amplitude of the variations seems to constantly increase from the X-rays up to the gamma-rays,
with the MeV-GeV range showing the largest variations along the electromagnetic spectrum.
On one side the correlation of the gamma-rays with the millimeter jet and with the optical flaring component \cite{jorstad09,chatterjee09}, 
and on the other side the lack of correlation between the hard X-rays and the millimeter emission \cite{soldi08} might indicate a different origin of the 
high-energy emission below and above the peak of the inverse Compton hump.

An extensive correlation study and an accurate modelling of the broad band spectral energy distribution during several epochs \cite{cutini09,abdo09c} will allow 
to shed more light on the origin of the gamma-rays in \thc\ and their connection to the emission in other bands.


\bigskip 

\end{document}